\documentclass{aa}
\usepackage{upgreek}
\usepackage{graphicx}
\usepackage[varg]{txfonts}

\usepackage{natbib}

\usepackage{soul}
\newlength{\linwx}
\usepackage{xcolor}
\usepackage{hyperref}
\usepackage{array}
\usepackage{amsmath}
\usepackage{multirow}
\usepackage{booktabs}

\hypersetup{colorlinks=true,linkcolor=purple,urlcolor=purple,citecolor = teal}

\graphicspath{ {./} }

\newcommand{\be}{\begin{equation}}
\newcommand{\ee}{\end{equation}}

\title{There is no disk mass budget problem of planet formation}

\author{Sofia Savvidou \inst{1}
\and Bertram Bitsch \inst{2,1}}

\offprints{S. Savvidou,\\ \email{savvidou@mpia.de}}

\institute{
Max-Planck-Institut f\"ur Astronomie, K\"onigstuhl 17, 69117 Heidelberg, Germany
\and
Department of Physics, University College Cork, Cork, Ireland}
\date{Received date / Accepted date }

\begin{document}

\abstract {The inferred dust masses from Class II protoplanetary disk observations are lower than or equal to the masses of the observed exoplanet systems. This poses the question of how planets form if their natal environments do not contain enough mass. This hypothesis has entered the literature as the "mass budget problem" of planet formation. We utilize numerical simulations of planet formation via pebble and gas accretion, including migration, in a viscously evolving protoplanetary disk, while tracing the time evolution of the dust mass. As expected, we find that the presence of a giant planet in the disk can influence the evolution of the disk itself and prevent rapid dust mass loss by trapping the dust outside its orbit. Early formation is crucial for giant planet formation, as we found in our previous work; therefore, our findings strengthen the hypothesis that planet formation has already occurred or is ongoing in Class II disks. Most importantly, we find that the optically thin dust mass significantly underestimates the total dust mass in the presence of a dust-trapping deep gap. We also show that the beam convolution would smear out the feature from a deep gap, especially if the planet forms in the inner disk. Such hidden dust mass, along with early planet formation, could be the answer to the hypothetical mass budget problem.}


\maketitle

\section{Introduction}

In the last decade, dust masses of disks have been estimated for several star-forming regions \citep[e.g.][]{Andrews+2013,Ansdell+2016,Barenfeld+2016,Pascucci+2016,Ansdell+2017,Cieza+2019,Cazzoletti+2019,Tychoniec+2018, Tychoniec+2020,Villenave+2021,Tobin+2020,vanTerwisga2022}, as well as some open clusters, \citep[e.g.][]{Eisner+2018,Ruiz-Rodriguez+2018}. However, an open question remains: how accurately do we measure the dust and gas mass budget of disks? If the answer is that we do have accurate measurements and we assume that Class II disks are the natal environments of forming planets, then a mass budget problem arises. The estimated disk masses are much lower than the inferred planetary system masses of observed exoplanets \citep{GreavesRice2010,NajitaKenyon2014,Mulders+2015,Mulders+2018}, implying that there is not enough material to form planets. 

Dust mass estimations in studies, such as the ones mentioned above, usually follow a flux-to-mass conversion initially proposed by \citet{Hildebrand1983}, which is therefore based on modeling and several assumptions. Among others, universal and constant opacity and temperature, optically thin emission, unknown chemical abundances or grain structure, and uncertain stellar ages are used, which would generally lead to underestimated masses. It is also assumed that most of the solid mass is in mm-sized grains \citep[see also][]{Miotello+2022arXiv,Manara+2022arXiv}. Alternative ways to measure the dust mass have been explored, for example through the dust lines \citep{Powell+2017,Powell+2019,Franceschi+2022} but these still have some limitations and lead to overestimated masses. 

Then, the conversion to total disk mass (gas+dust) is usually done considering an interstellar medium- or solar-like dust-to-gas ratio of 1\% to $\sim$1.5\%. However, several studies find that the dust-to-gas ratio can be higher than 1\% \citep[e.g.][]{Appelgren+2020,Kama+2020}. Especially in regions with significant dust trapping, it is unreasonable to assume that any initial dust-to-gas ratio will remain relevant. 

Recently, there has been increasing work contemplating whether these limitations of the dust mass estimates could lead to significantly underestimated masses \citep{Galvan-Madrid+2018,BalleringEisner2019,Zhu+2019,Bergez-Casalou+2022}. \citet{Manara+2018} suggest that another possible way out of this problem could be early planet formation so that it is already ongoing in observations of early protoplanetary disk evolutionary stages (younger than 3 Myr). 

There is increasing evidence that this is the case; for instance, \citet{Harsono+2018} reported dust growth to millimeters for a very young disk (around 0.1 Myr) around a solar-type protostar, while \citet{SheehanEisner2018,Segura-Cox+2020}, and \citet{Cieza+2021} find disk substructures already in very young disks (younger than 1 Myr) which is not necessarily expected by the formation of the disk. Such observed substructures in disks (e.g. gaps or rings) can be theoretically expected outcomes of the formation processes for planets with high enough masses. There are also meteorological analyses confirming that some of the building blocks of the solar system have formed within 1 Myr \citep{Kruijer+2017,Wadhwa+2020}. These observations all point to dust growth being very efficient even as the disk itself still forms and would, thus, allow for early planet formation. 

From a theoretical point of view, this possibility is also supported by the pebble accretion scenario, because it can provide high pebble accretion rates at the earliest stages of evolution \citep{Bitsch+2015b,Johansen+2019,TanakaTsukamoto2019}. Additionally, \citet{Appelgren+2020} find that the observed low disk dust masses at later evolutionary stages can be consistent with even very massive young disks that experience rapid radial dust drift. 

In our previous work \citep{SavvidouBitsch2023}, we performed numerical simulations of planet formation through pebble and gas accretion, including migration, in viscously evolving protoplanetary disks, investigating how the disk conditions influence the resulting planetary masses. In this work, we utilize the same set of simulations, and discuss how the dust evolves during the planet formation process and what this implies for the hypothesis of the disk mass budget problem. 

\section{Methods}
\label{Sec:Methods}

\subsection{Model and parameters used}
The model was described in detail in \citet{SavvidouBitsch2023}. Briefly, the numerical simulations of planet formation in a protoplanetary disk include pebble growth and drift \citep{Birnstiel+2012}, pebble evaporation and condensation at ice lines \citep{SchneiderBitsch2021a}, planet growth via pebble \citep{JohansenLambrechts2017} and gas accretion \citep{Ndugu+2021} as well as planet migration \citep{Paardekooper+2011}. The initial planetary mass of the embryos is determined by the pebble transition mass, at which the planet starts efficient accretion from the Hill regime \citep{LambrechtsJohansen2012}. For these simulations, we used the 1D semi-analytic \texttt{chemcomp} code \citep{SchneiderBitsch2021a}. We note that we do not include planetesimal formation and we do not include dust filtering across gaps. An extended discussion about the limitations of our models is in Appendix \ref{App:Caveats}.

Each simulation contained one growing planet, and we assumed that the star was of solar mass. We performed a parameter study where the $\alpha$-viscosity parameter ranges from 0.0001 to 0.001, the initial disk mass ranges from 0.01 to 0.1 M$_{\odot}$, the initial disk radius ranges from 50 to 200 AU, the dust fragmentation velocity ranges from 1 to 10 m/s, and the planetary embryos are placed in the disk from 1 to 50 AU, with starting times from 0.1 to 1.3 Myr \citep[Table \ref{Tab:parameters}, see also][]{SavvidouBitsch2023}. Our whole sample then is all possible combinations of these parameters (first six in Table \ref{Tab:parameters}) and two different values of the dust-to-gas ratio (78400 runs). The standard set of parameters used to discuss example cases is in bold in Table \ref{Tab:parameters}. 

\subsection{Optically thin dust mass estimates}

The continuum flux of the dust emission obtained from observations is converted to a dust mass via the \citet{Hildebrand1983} approximation
\be \label{Eq:Hildebrand} F_\nu=\frac{B_\nu(\overline{T_d})\overline{\kappa_d}}{d^2}M_d~, \ee
assuming that the emission is optically thin and is well described by an average dust temperature $\overline{T_d}$ \citep{Andrews+2005} and an average opacity $\overline{\kappa_d}$ \citep{Beckwith+1990}. 
In the Rayleigh-Jeans approximation (relevant for large wavelengths), the Planck spectrum is written
\be \label{Eq:Rayleigh-Jeans} B_\lambda(T)=\frac{2 c k_BT}{\lambda^4}~. \ee
The intensity (for an optically thin emission) is written as 
\be I_\lambda=B_\lambda(T_d)\tau_\lambda~,\ee
where $\tau_{\lambda}$ is the optical depth. In the Rayleigh-Jeans limit the intensity is then
\begin{equation} 
\label{Eq:Intensity}
I_\lambda=\frac{2 c k_B\overline{T_d}}{\lambda^4}\overline{\kappa_d}\Sigma_{d,\tau_\lambda<1}~.
\end{equation} 
The surface density $\Sigma_{d,\tau_\lambda<1}$ corresponds to the disk regions in our models where the disk is optically thin ($\tau_\lambda$<1).
We used the flux-to-mass approximation of Eq. \ref{Eq:Hildebrand} to estimate the disk dust masses from our models with the same assumptions as in the observed sources and to compare them with the actual dust masses of the modeled disks. 

\section{Dust mass evolution during planet formation}
\label{Sec:Dust_masses}

\subsection{Example cases}
\label{subsec:Examples}

\begin{figure}
\centering
\includegraphics[width=.85\columnwidth]{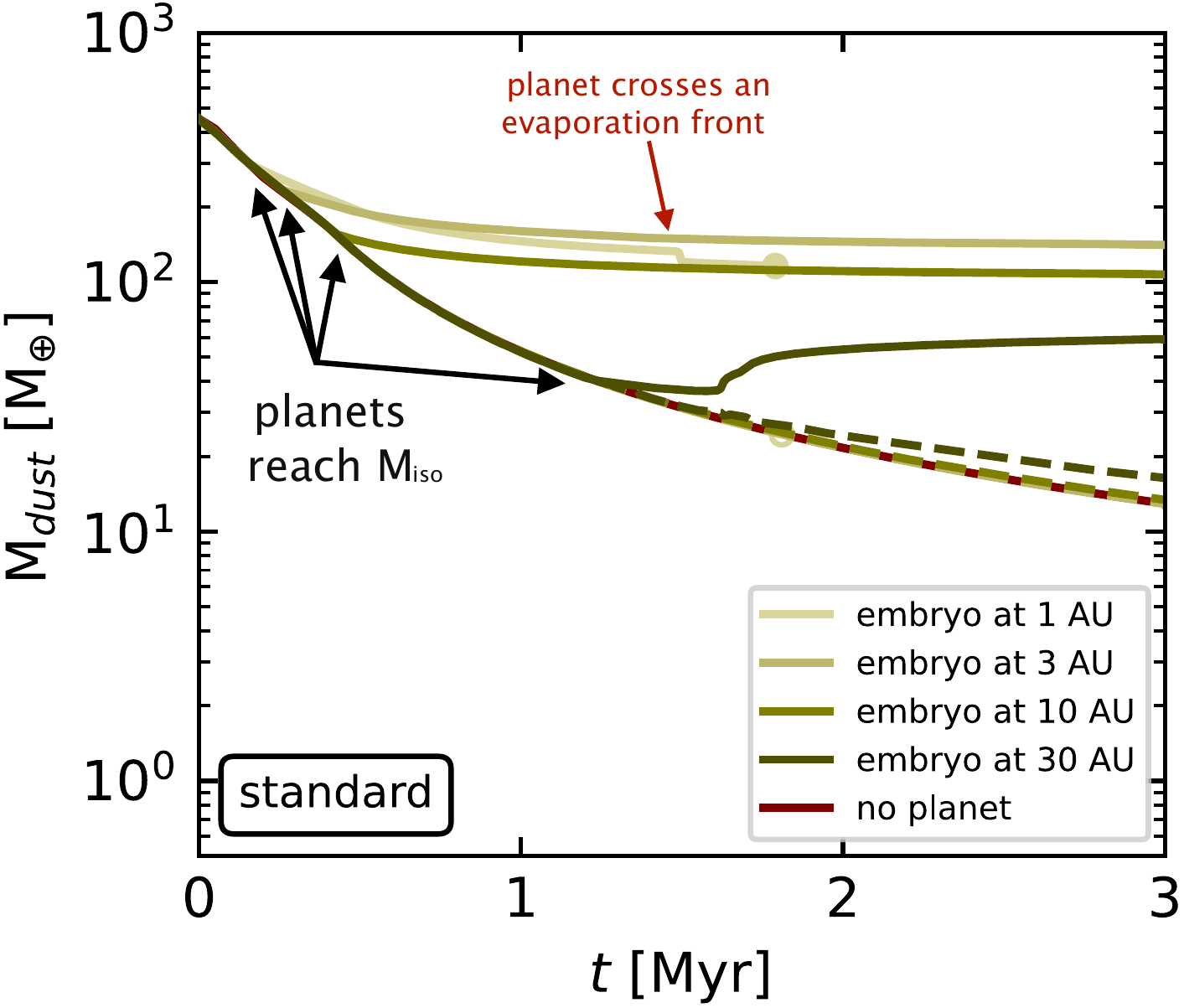}
\caption{Dust mass as a function of time for the same planets as in Fig. 1 in \citet{SavvidouBitsch2023}, using the standard set of parameters (Table \ref{Tab:parameters}). The dashed lines show the time evolution of the optically thin dust mass and the red line shows the time evolution of the dust mass for a simulation without a planet. The dot marks the last dust mass if a planet reaches the inner edge of the disk.}
\label{Fig:t-Mdust_main}
\end{figure}

\begin{figure*}
\centering
\includegraphics[width=\textwidth]{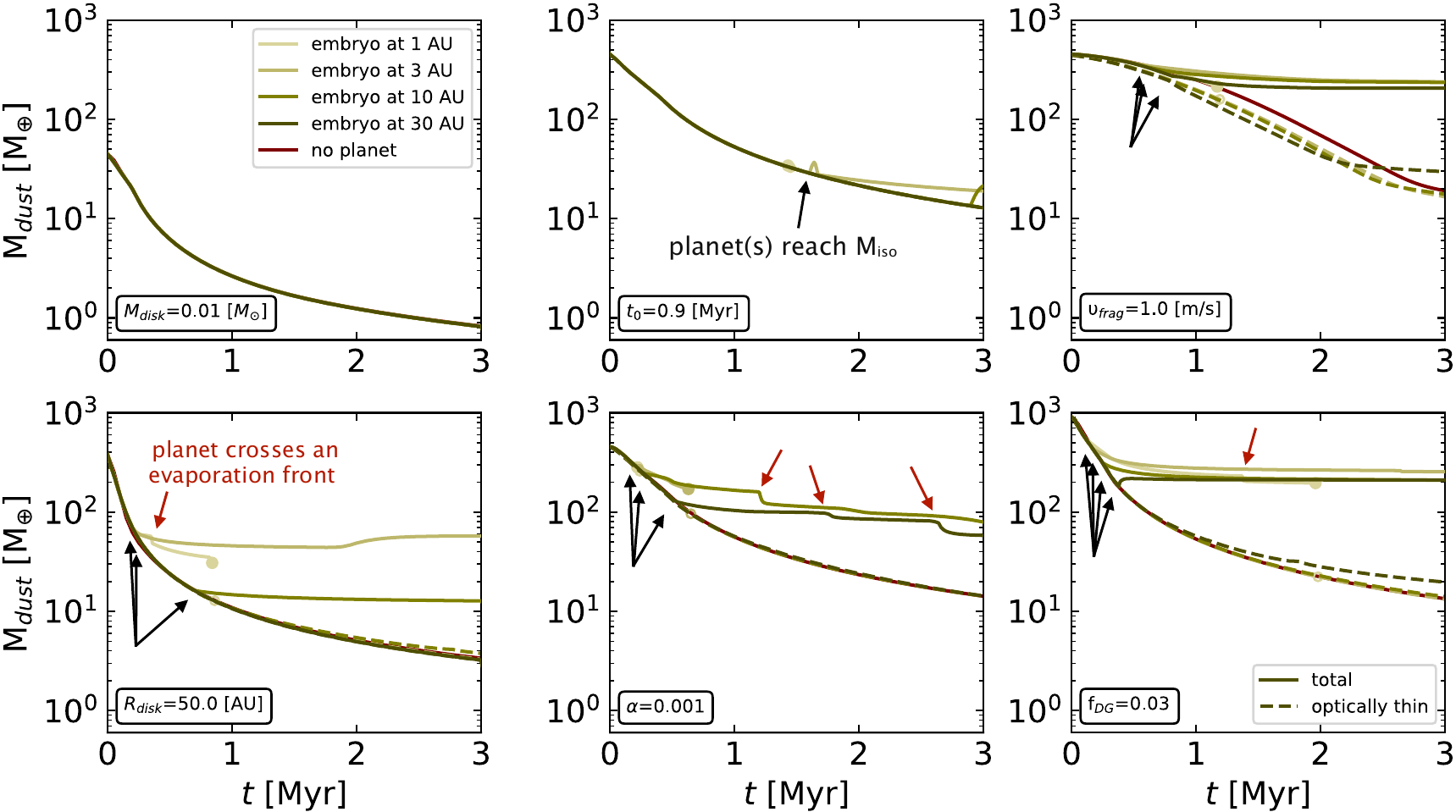}
\caption{Same as Fig. \ref{Fig:t-Mdust_main}, with one parameter changed (marked in each panel) compared to the standard model (bold in Table \ref{Tab:parameters}). These correspond to the same planets as in Fig. 2 in \citet{SavvidouBitsch2023}.}
\label{Fig:t-Mdust}
\end{figure*}

In Fig. \ref{Fig:t-Mdust_main}, we show the time evolution of the dust mass for four embryos that started growing at 1, 3, 10, and 30 AU (solid green lines), along with a simulation without a planet (solid red line). The examples shown here are the same simulations as in Fig. 1 in \citet{SavvidouBitsch2023}. We also estimate the dust mass according to Eq. \ref{Eq:Hildebrand}, using the surface density in the optically thin regions only (dashed green lines). Some of the planets (mainly the ones originating from 1 AU-see Figs. 1-2 in \citet{SavvidouBitsch2023}) reach the inner edge of the disks in our models long before the gas dissipates. In this case, we no longer plot the dust evolution and we indicate with circles the dust mass at the time when they reach the inner edge. The solid circles correspond to the total dust mass, and the hollow ones correspond to the optically thin dust mass.

Without any planet growing, the dust decreases as it drifts inwards and evaporates at the icelines or is lost by stellar accretion. The dust mass evolution in our models where planets are growing is heavily dependent on the growth evolution of the planet. The embryos originating at 1, 3, and 10 AU reach the pebble isolation mass within 0.5 Myr (shown in Fig. 1 in \citet{SavvidouBitsch2023}), and begin effectively blocking the drifting dust exterior to their orbit (marked with the arrow in Fig. \ref{Fig:t-Mdust_main}). Therefore, the dust content in these disks almost plateaus beyond 0.5 Myr. The amount of dust blocked in our models depends on several things, such as the location of the embryo and how long it took until the planet reached the pebble isolation mass and trapped the dust. For example, the planet originating at 10 AU reaches the pebble isolation mass earlier and there is more dust mass exterior to its orbit compared to the one originating at 30 AU, therefore more dust gets trapped. 

In the disk with the planet originating at 30 AU, the dust mass increases shortly after 1.5 Myr. Initially, the pebbles drift inwards and evaporate; thus, the dust mass decreases. Then the planet starts growing and opens a gap; however, the position of the planet is such that the pressure bump generated by the opening gap is close to an evaporation front. As the planet becomes more massive, the gap it opens becomes wider, and therefore the peak of the pressure bump moves outwards. Then some amount of volatiles diffuses over the evaporation front again, recondenses and increases the dust mass. 

The optically thin dust masses for all of the models (dashed green lines) evolve similarly to the dust mass in a disk without a planet, regardless of the initial location of the embryo, its orbital evolution, and its final mass. The difference between the total dust mass and the optically thin dust mass is almost one order of magnitude. Even though giant planets generate a pressure bump, where many particles can be trapped, the optically thin approximation underestimates the real dust mass.

In Fig. \ref{Fig:t-Mdust}, we only change one parameter at a time, as in Fig. 2 in \citet{SavvidouBitsch2023}. In that work, we showed that no giant planet forms when the initial disk mass is low (top, left plot), due to the lack of enough solid material to build a massive core, or when the embryo is injected too late (top, middle plot). Without a giant planet to block the dust, it drifts and decreases even within 0.5 Myr. The evolution is, hence, the same under the optically thin dust assumption and for the model without a planet. The same is true for the planet originating at 30 AU in the disk with a radius of 50 AU (bottom, left plot) that did not reach the pebble isolation mass. In the rest of the examples for this case, the dust mass stops decreasing after the planet opens a massive gap in the disk.

As a giant planet moves, it carries the pressure bump with it (and thus the pebbles). Once it crosses an evaporation front, pebbles evaporate and there is a ``dip" (sharp decrease) in the total dust mass (e.g., in Fig. \ref{Fig:t-Mdust_main}, for the disk with a planet originating at 1 AU, or in the middle, left plot of Fig. \ref{Fig:t-Mdust} for the planet that originated at 1 AU (marked with red arrows)). These ``dips" are proportional to the amount of material that evaporates, so larger ``dips" are expected if the planet crosses evaporation fronts corresponding to more abundant species. Additionally, the ``dips" are more numerous for $\alpha$=0.001, because the planets migrate faster in that case and cross multiple evaporation fronts. 

We, overall, find that when a massive core forms and opens a gap in the disk, the remaining dust mass in the disk is significantly higher compared to the one estimated with the optically thin dust mass assumption. The sooner the gap opens, the larger the difference is between the actual total dust mass near the end of the disk lifetime and the optically thin estimate. In most cases, this difference is at least one order of magnitude.

\subsection{Radial intensity profiles}

\begin{figure*}
\centering
\includegraphics[width=\textwidth]{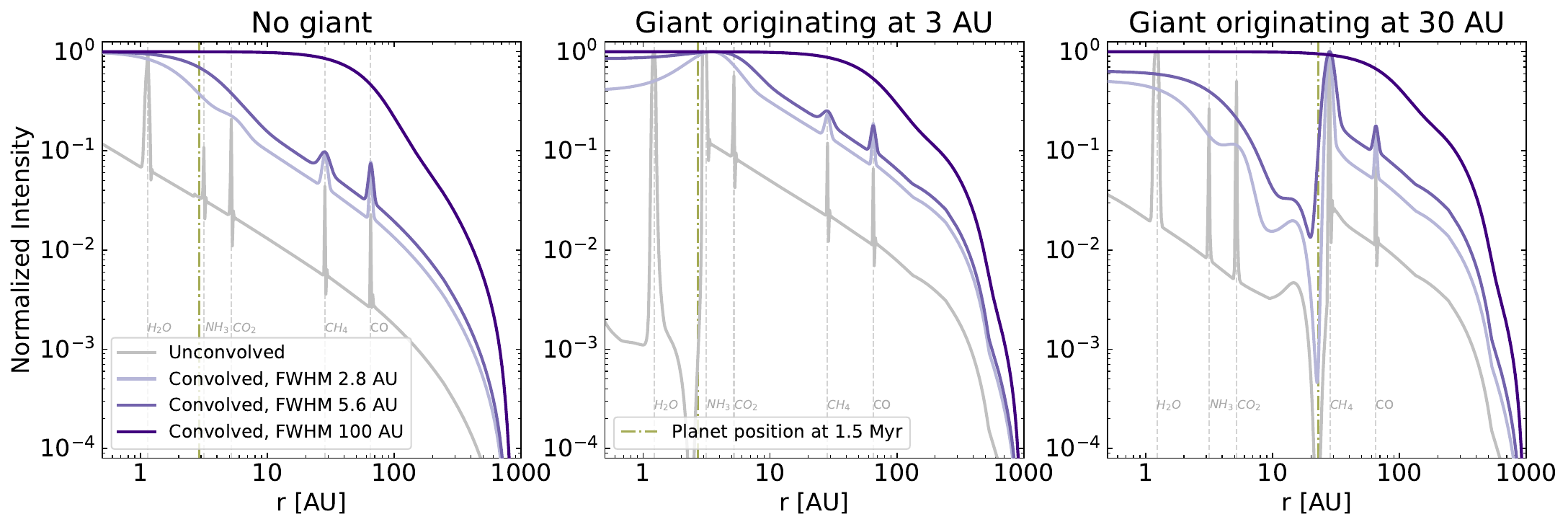}
\caption{Normalized intensity as a function of orbital distance at 1.5 Myr, comparing the uncolvolved intensity with the ones convolved with three different beams of 0.02" and 0.04" at 140 pc (2.8 and 5.6 AU), and 0.34" at 232 pc (100 AU). The gray dashed lines show the evaporation fronts that mainly cause the spikes in the intensity. The examples with a giant correspond to the standard set of parameters (Fig. \ref{Fig:t-Mdust_main}), while the example without a giant corresponds to a low-mass disk (top, left plot in Fig. \ref{Fig:t-Mdust}).}
\label{Fig:r-I}
\end{figure*}

It is interesting to consider whether the giant planets that we discussed in the examples of the previous section would lead to observable features at the disk. To this end, in Fig. \ref{Fig:r-I}, we show the normalized intensity of the dust emission in our models as a function of the orbital distance. The intensity is normalized to the peak intensity along the radius of the disk. For these calculations, we use the Rayleigh-Jeans approximation, following Eq. \ref{Eq:Intensity}.

For simplicity, we summarize these examples with three distinct cases, at 1.5 Myr after the disk starts evolving. In the left plot of Fig. \ref{Fig:r-I}, we show a disk in which no giant formed (example shown in the top, left plot in Fig. \ref{Fig:t-Mdust}), in the middle plot, one where the giant originated at 3 AU (shown in Fig. \ref{Fig:t-Mdust_main}), and in the right plot one where the giant originated at 30 AU (shown in Fig. \ref{Fig:t-Mdust_main}). 

The giants cause a deep gap in the disk and, therefore, also in the intensity. However, the question is if these features would survive the convolution with a beam. We overplot in the same figure the normalized radial intensity convolved with three different beams. The FWHM of the two beams are 0.02" and 0.04", corresponding to 2.8 and 5.6 AU assuming a source at a distance of 140 Pc. These are chosen to represent some of the observations with the highest resolution so far \citep[e.g.][]{Benisty+2021,Andrews+2018}. We also show the convolved intensity with a beam of 0.34", that corresponds to 100 AU at the distance of the Perseus molecular cloud \citep[an average of the sources discussed in][]{Tychoniec+2020}. 

The gap caused by a giant that started forming at the inner disk would be missed in observations even with high resolution (FWHM of a few AU) and with insufficient resolution even at large distances. At the same time, the common methods for estimating the disk dust mass could ``hide'' the trapped mass at the planetary-induced pressure bump, as we discussed before. Both of these could be part of the main reasons why we estimate such low disk masses from observations that appear to be limiting for planet, let alone giant planet, formation.

\subsection{Whole sample}

\begin{figure}[b]
\centering
\includegraphics[width=.85\columnwidth]{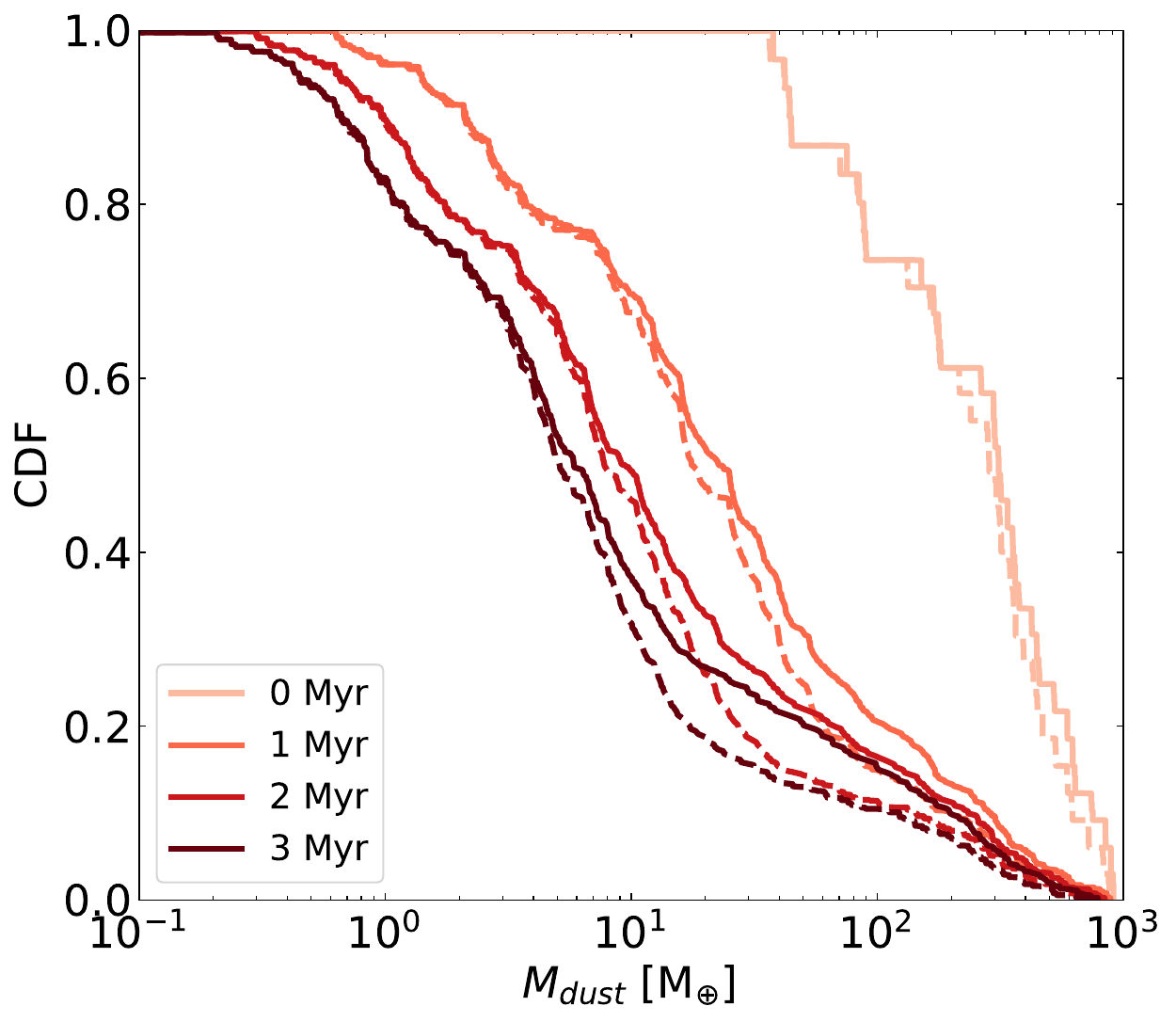}
\caption{Cumulative distribution functions for the disk dust mass of our models at different times (0-3 Myr). The dashed lines correspond to the dust mass in the optically thin limit.}
\label{Fig:CDF-Mdust}
\end{figure}

We generalize, now, our results to our whole sample and present, in Fig. \ref{Fig:CDF-Mdust}, the CDFs of the dust mass at different times, from the beginning of the simulations, until 3 Myr (near the end of the lifetime of the disk in our models), every 1 Myr. We note that we excluded here the simulations where the planets reach the inner edge of the disk. The evolution of them after this remains highly unknown and hence we consider uncertain how they will influence the dust evolution. 

We find that the difference between the total dust mass (solid lines) and the optically thin dust mass (dashed lines) increases with time. If there is nothing to prevent radial drift, the dust decreases in time, and this is reflected in the optically thin dust mass estimates. As we show in Figs. \ref{Fig:t-Mdust_main} and \ref{Fig:t-Mdust}, the time evolution of the optically thin dust mass closely resembles that of the models without any dust trapping.

Interestingly, the time evolution from our models resembles the time evolution that we see when plotting the CDFs of observed disks grouped by classes \citep[see for example Fig. 2 in][]{Drazkowska+2022arXiv} and specifically focusing on Class 0 to Class II objects. A direct comparison with the observed dust mass estimates is not trivial and was not intended in this work; however, we also note that the CDFs at the end of our simulations resembles the ones for Class II disks. In other words, the most similar CDF from our models to the observed Class II disks is the one where planet formation has already happened, especially accounting for the optically thin dust mass estimates of our models. 

This strengthens the hypothesis that planet formation starts early and has already happened or has been ongoing for at least a few million years in Class II systems that were commonly assumed to be the progenitors of planets. As expected, we show a mass loss in time, similarly to the CDFs of observed disks in ascending evolutionary stages. Even though the dust mass decreases significantly over time - especially if accounting for the optically thin dust mass - this poses no problem for planet formation because it has already happened.

We, hence, want to point out that the optically thin dust emission assumption for observations could be leading to a significant underestimation of the dust masses. The trapped dust due to a giant planet-induced pressure bump could be unresolved and thus not accounted for accurately. This also highly depends on the location of the giant planet, as the inner regions are generally more difficult to resolve but seem to be more favorable for giant planet formation \citep{SavvidouBitsch2023}. This conclusion, though, also be the same for any mechanism that could cause dust trapping.

\section{The mass budget problem}
\label{Sec:Mass_budget_problem}

The last decade's burst of millimeter disk surveys, in several star-forming regions, has led to extensive total mass estimations which appear to be not high enough to explain planet formation, especially without 100\% efficiency in the formation mechanisms \citep{GreavesRice2010,NajitaKenyon2014,Lambrechts+2014,Manara+2018,Bitsch+2019}. Several solutions to this hypothetical problem have been discussed, but two categories seem to be the most plausible. On one hand, there could be an underestimation of the masses caused by the assumptions in the flux-to-mass conversion or contributions from optically thick regions. On the other hand, Class II disks could mainly contain the leftovers from planet formation if it starts much earlier than previously assumed. 

The flux-to-mass conversion (Eq. \ref{Eq:Hildebrand}) relies by default on a mostly optically thin emission for the dust at (sub-)millimeter wavelengths (similar when using the Rayleigh-Jeans approximation). However, the observed disks could be optically thick if dust scattering is considered \citep{Zhu+2019,Dullemond+2018} or partially optically thick in smaller regions (e.g. rings) \citep{Tripathi+2017,Andrews+2018}. \citet{Tychoniec+2020} estimate the disk masses in the Perseus star-forming region, comparing ALMA (1.1-1.3 mm) and VLA (9 mm) data, as well as dust mass estimates from other regions. 
Their estimated dust mass medians for Class 0 and I disks from VLA are higher compared to the estimates from the ALMA observations, but not for the Class II disks of their sample, so this could be pointing towards an early formation of planets and a significant amount of optically thick dust at ALMA wavelengths in the very young disks. \citet{Xin+2023}, perform SED modeling to constrain the disk dust masses in Lupus and find $\sim$1.5-6 times higher mass estimates compared to the ones estimated via \citet{Hildebrand1983}.

Even if the millimeter emission is indeed optically thin (especially for Class II disks), the terms that go into Eq. \ref{Eq:Hildebrand} all have great uncertainties and are usually oversimplified. For example, constant values are commonly used for the temperature \citep{AndrewsWilliams2005} or the opacity \citep{Beckwith+1990,Andrews+2013}, regardless of the disk or population properties, or the evolutionary stage of the disk. As an example, \citet{Pinilla+2022} find that if the outer disk is as cold as the interstellar medium, then the estimated optical depth of observed disks could be well above unity. The disk radius can, also, be difficult to constrain and this can lead to a significant underestimation of the dust masses \citep{Liu+2022}. Then the dust-to-gas ratio to convert the dust masses into total disk masses is also an educated guess and can deviate from the interstellar value \citep{WilliamsCieza2011}.

The other possible explanation for the discrepancy between the masses of the observed exoplanets and the observed protoplanetary disks could be that planet formation starts in Class 0/I disks, in which the estimated mass seems to be sufficient \citep[e.g.][]{Tychoniec+2018}. The observed decrease in the masses of the Class II disks in different star-forming regions can be explained by the radial drift of the dust \citep{Appelgren+2020}, the formation of larger bodies that are observationally unattainable, and/or the entrapment of solids into some region due to pressure bumps \citep{Pinilla+2020}. Such dust traps, whether caused by giant planets or not, can lead to optically thick substructures that are not necessarily accounted for in the dust mass estimations, as we find in this work \citep[see also][]{Miotello+2022arXiv,Manara+2022arXiv}.

The observed substructures even in young disks could be considered an evolutionary signature, supporting early giant planet formation that creates dust traps and observable features. They are mainly found in massive disks \citep[see also][]{Drazkowska+2022arXiv,Bae+2022arXiv}, that are also more probable to host giant planets \citep{SavvidouBitsch2023}. Such disks have often remained massive even at older ages, indicating that giants could indeed be the creators of long-lived pressure bumps that trap dust and give structure to the disks \citep{vanderMarelMulders2021}. The apparent lack of substructures in observations of less massive disks could be explained by insufficient resolution \citep{Bae+2022arXiv} or by the loss of information from the chosen deconvolution technique \citep[see for example][]{Jennings+2022}. 

The brightness of the substructures in the mm-continuum could be related to the timing of their formation \citep{Garate+2023}, which would help understand their origin. However, there are degeneracies that make it difficult to link the observed substructures to a planet and its properties \citep{Bergez-Casalou+2022,Garate+2023,Tzouvanou+2023}, potentially explaining the lack of direct planet detections even in transition disks.

\section{Conclusions}
\label{Sec:Conclusions}

We conclude that the mass budget problem does not exist, based on the following: 
\begin{enumerate}
\item The assumptions concerning the dust mass estimates can at the very best lead to a lower limit of the dust mass and cannot accurately estimate the total disk mass.
\item We concluded in our previous work \citep{SavvidouBitsch2023}, considering also the exoplanet occurrence rates, that early planet formation is crucial, especially in regards to giant planet formation. 
\item Trapping dust in the disk (via any mechanism) will lead to a significant underestimation of the solid mass under the common assumptions used in the modeling of disk observations (i.e. the flux-to-mass approximation under the optically thin emission assumption).
\item Giant planets are expected to create a pressure bump exterior to their orbit that will trap the inward drifting dust (pebbles) at least in the short-term, potentially depending also on the Stokes numbers of the dust grains, as well as the viscosity of the disk. Any trapped dust will be largely unaccounted for by the optically thin emission approximation. Additionally, planet formation has already happened in this case which means that there should be no missing dust budget dilemma.
\end{enumerate}

We should aim for a combination of dust mass estimates from multiwavelength observations whenever possible since dust emission at longer wavelengths is more probable to be optically thin. We should also consider whether the current approaches (e.g. algorithms or observational durations) sufficiently resolve the sources and whether some information could still be elusive, especially at the inner regions of disks or in less massive/bright sources. We conclude that the mass budget problem hypothesis is not an issue and it is thus not necessary to have 100\% efficiency for planet formation.

\begin{acknowledgements}

B.B. and S.S. thank the European Research Council (ERC Starting Grant 757448-PAMDORA) for their financial support. S.S was a Fellow of the International Max Planck Research School for Astronomy and Cosmic Physics at the University of Heidelberg (IMPRS-HD) during part of the work. S.S. thanks Cornellis P. Dullemond and Nicol\'as Kurtovic for discussions that helped with one of the paper's subsections.                        
\end{acknowledgements}

\bibliographystyle{aa}
\bibliography{SavvidouBitsch24}

\begin{appendix}
\section{Parameter study}

\begin{table}[h]
\centering
\begin{tabular}{@{}ccc@{}}
\toprule[1.2pt]
Parameter               & Values               &                              \\ \midrule
$M_0$ {[}$M_{\odot}${]}  & 0.01, 0.04, 0.07, {\bf 0.1}   & initial disk mass            \\ \addlinespace
$R_0$ {[}$R_{\odot}${]}  & 50, 100, 150, {\bf 200}   & initial disk radius          \\ \addlinespace
$\alpha$                & {\bf 0.0001}, 0.0005, 0.001 & $\alpha$-viscosity parameter \\ \addlinespace
$t_0$ {[}Myr{]}          & {\bf 0.1}, 0.5, 0.9, 1.3   & starting time of embryo      \\ \addlinespace
$\alpha_{p,0}$ {[}AU{]} & 1-50 every 1  & initial position of embryo   \\ \addlinespace
$u_{frag}$ {[}m/s{]}    & 1, 4, 7, {\bf 10}    & fragmentation velocity     \\ \addlinespace
$f_{DG}$                & 0.01, {\bf 0.015}, 0.03 & dust-to-gas ratio    \\ \bottomrule[1.2pt]
\end{tabular}
\caption{Parameters used in the simulations for Fig. \ref{Fig:CDF-Mdust}. We mark in bold the standard set, which is used as a reference in Fig. \ref{Fig:t-Mdust_main}-\ref{Fig:r-I}.}
\label{Tab:parameters}
\end{table}

\section{Caveats}
\label{App:Caveats}

\subsection{Planetesimal formation and accretion}

In our work here, we did not take planetesimal formation and accretion into account. First of all, the planetesimal formation efficiency still depends on several (unknown) parameters, for example how efficiently pebbles are transformed into planetesimals \citep[see][]{DrazkowskaAlibert2017,Lenz+2020} or which their preferred formation locations are \citep{Andama+2024,SchoonenbergOrmel2017}. \citet{Danti+2023} show that the important differences between planetesimal and pebble accretion models lie in the composition of the planets that are forming, while the final planetary masses remain similar, especially at lower viscosities. Additionally, the ability to form giant planets seems to remain consistent with or without planetesimal accretion \citep{KesslerAlibert2023}. These were the main discussion points of our previous paper and the basis for this paper's work.

The accretion of planetesimals formed at gap edges exterior to giant planets also does not result in a significant change of the planetary mass, due to the inefficiency of planetesimal accretion during the gas phase, even if 100\% of the dust is transformed into planetesimals \citep{Eriksson+2022}. This indicates that these planetesimals do not change the growth of the forming planet. 

Nevertheless, it is expected that if planetesimal formation was included then some part of the dust mass that we show in our plots here would be used to form planetesimals. Therefore, the total dust mass (solid lines) in Fig. \ref{Fig:CDF-Mdust} can be considered the maximum solids mass that the disks can retain. The solids mass that would be in planetesimals would then not be observable. The formation of such larger bodies (whether pebbles greater that a few centimeters or boulders/planetesimals) can contribute to the low disk mass measurements, while there is still no problem to planet formation, as planetesimal formation mostly influences the inner disk chemistry, but not the growth of the planets themselves \citep{Danti+2023,KesslerAlibert2023}.

It is important to consider planetesimal formation and accretion in future work that discusses the compositional evolution of the disk or the forming planets. In the present work, though, our conclusions regarding the trapped dust mass are based on the final planetary masses and the formation of giant planets, both of which would be minimally influenced by planetesimal formation and accretion, as we discuss above. Finally, the measured protoplanetary disk dust masses pose no constraint on planet formation regardless because the planets have already formed in the observed disk population.

\subsection{Dust filtering through gaps}

In our work here, we assumed that all dust grain sizes can be efficiently trapped at the planetary-induced pressure bump. Recent studies suggest that some filtering of the dust grains might be taking place in such cases; trapping the larger grains but allowing the smaller dust to follow the gas motion and pass through the gap, perhaps enhanced by dust fragmentation \citep{Stammler+2023}. However, this significantly depends on the planetary mass, the strength of the dust diffusivity, and the core growth timescale (which again depends on the disk conditions). 

It is, also, important to consider that as dust grains approach a deep gap, where the gas density decreases sharply, their Stokes numbers will rapidly increase; therefore, they will still decouple from the gas and remain trapped \citep{Weber+2018}. This will lead to a varying location where particles of different sizes will be trapped, and thus potentially a wider bump but the dust will remain mostly concentrated exterior to the gap. Additionally, the aforementioned studies do not include planetary migration. The comparison between the planetary migration speed and the velocity of the dust grains (that are well-coupled to the gas if small enough) can be a limiting factor to the permeability of a planetary gap \citep{Morbidelli+2023}. 

Even with some degree of dust filtering through the gap, the millimeter fluxes in disks that include gaps/bumps will be lower compared to a smooth disk \citep{Pinilla+2021}. This aligns with our conclusions here that the presence of a deep gap in the disk could affect the disk structure but this is hard, if not impossible, to detect observationally with the current technological means and the common methods.

Let us instead consider that regardless of the above, the dust indeed manages to be very efficiently transported through the gap. Then the dust mass would decrease faster and would approach the fractions shown in our models without a planet. But then our conclusion remains the same: There is still no mass budget problem because planet formation has successfully happened and our measurement of the disk dust masses cannot be limiting to planet formation. 

\subsection{1D vs 2D and two-poppy vs full grain size distribution}

We used here a 1D semi-analytic code to simulate planet formation via pebble accretion and explore a wide range of parameters to investigate how the initial conditions in the disk influence planet formation and how this influences the connection to the disk observations. The 1D approach allows for fast and thus numerous calculations, however 2D (or 3D) and full hydrodynamical models are expected to offer more accurate and detailed physics. We also utilize here the two-population approach for the dust distribution rather than a full grain size distribution or even a full coagulation-fragmentation model. This approach allows for significantly faster calculations that again allow for a wide exploration of the parameter space that influences planet formation. 

In relation to our results, we expect for example that the spikes in the intensity profile would be wider and potentially less strong in a full coagulation 2D model because the asymmetric effects that a planet pressure bump would induce would be ``smeared out" \citep{Drazkowska+2019}. This would only make the presence of a deep gap in the disk even harder to detect through observations (see Fig. \ref{Fig:r-I}).

\end{appendix}

\end{document}